\documentclass[
 reprint,
superscriptaddress
]{revtex4-2}
\usepackage{multirow}
\usepackage{ragged2e}
\usepackage{graphicx}
\usepackage{bm}
\usepackage{siunitx}
\usepackage{acro}
\DeclareAcronym{MLIP}{short=MLIP, long=Machine Learning Interatomic Potential}
\preprint{APS/123-QED}

\begin{document}

\title{Kink-Helium Interactions in Tungsten: \\ Increased Dislocation Mobility in the Infinitely Dilute Regime}%

\author{Matthew Nutter}
\email{m.nutter@warwick.ac.uk}

\affiliation{Department of Physics, University of Warwick, Coventry CV4 7AL, United Kingdom}
\affiliation{Warwick Centre for Predictive Modelling, School of Engineering, University of Warwick, Coventry CV4 7AL, United Kingdom}
\author{James R. Kermode}%
\affiliation{Warwick Centre for Predictive Modelling, School of Engineering, University of Warwick, Coventry CV4 7AL, United Kingdom}
\author{Albert P. Bartók}
\affiliation{Department of Physics, University of Warwick, Coventry CV4 7AL, United Kingdom}
\affiliation{Warwick Centre for Predictive Modelling, School of Engineering, University of Warwick, Coventry CV4 7AL, United Kingdom}

\begin{abstract}

Point defects such as interstitial atoms are known to be attracted to screw dislocations. Understanding these interaction mechanisms is key to predicting the plasticity of real materials. Using a new machine learning interatomic potential derived from \textit{ab initio} calculations of helium in tungsten, we investigate the binding of small helium clusters ($\mathrm{He}_{n}$, ${n}=1-3$) to screw dislocation kinks. We find that helium binds significantly more strongly to kinks than to straight dislocation segments. For a single helium atom, the preference reduces the kink pair nucleation energy from 1.58~eV in pure tungsten to 0.48~eV when helium binds to the vacancy kink, indicative of increased screw dislocation mobility (material softening). In the case of ${n}=2$, kink binding stabilises the kink pair configuration as the ground state, while the straight dislocation is metastable; the two are separated by a 0.60~eV barrier that is the rate determining step for dislocation motion. The energy difference between the ground state kink pair and the metastable straight dislocation increases for ${n}=3$, connected by a 1.00~eV barrier, indicating a progressive loss of the softening effect with increasing cluster size. Molecular dynamics simulations at 900~K support the proposed existence of helium-induced softening in this dilute regime.

\end{abstract}

\maketitle
\onecolumngrid
\vspace{-2.75\baselineskip}
\begin{center}
    \textcopyright{} 2026 British Crown Owned Copyright / AWE
\end{center}
\vspace{-0.25\baselineskip} 
\twocolumngrid

\section{Introduction}

Tungsten is a primary candidate for the plasma-facing armour components of future nuclear fusion reactors, due to its high melting point, resistance to radiation damage, and thermal conductivity \cite{Rieth2013-uz}. Helium atoms will be present from the fusion process and transmutation reactions. Such defects are believed to interfere with the movement of screw dislocations with $\frac12\langle111\rangle$ Burgers vector, $\it{b}$, which typically propagate through the crystal by thermally activated nucleation and migration of kink pairs.
Since the plasticity of body-centred cubic (bcc) metals such as tungsten is largely determined by the movement of these dislocations, it is crucial to have a detailed understanding of how they behave in the presence of impurities to ensure the durability of these components. 

Predictions indicate that after five years, reactor components may have helium concentrations of up to 20~appm \cite{Gilbert2012-bp}, a level that has been demonstrated in experiments to suppress recrystallisation and increase the ultimate tensile stress \cite{Hasegawa2021-cq}. 
Thermodynamic modelling indicates that a helium concentration on the order of 1~appm corresponds to a regime in which helium is expected to be abundant and strongly localised at extended defects such as dislocations \cite{Grigorev2023-fm}. Helium interstitials are also known to attract one another and form clusters in bulk tungsten \cite{Becquart2006-wk}, motivating consideration of both individual helium atoms and small helium clusters. At substantially higher concentrations, nanoindentation experiments on tungsten single crystals implanted with 3000~appm helium suggest that helium-induced defects act as obstacles to dislocation motion \cite{Das2018-mh}. At a still higher concentration of 200,000~appm (20 atomic \%), transmission electron microscopy has directly visualised reduced screw dislocation mobility \cite{Zheng2021-bn}.

The present work instead examines the interaction of small helium clusters with a $\frac12\langle111\rangle$ screw dislocation in the infinitely dilute, single-cluster limit. By this, we mean that only one helium cluster is present on the dislocation line: at most one of the two kinks forming a kink pair can be bound to helium, while the other remains unbound and can migrate without encountering a second cluster. 

We investigate the possibility of helium-induced softening in this regime, motivated by the observation that hydrogen in bcc metals exhibits a concentration-dependent transition from softening at low concentrations to hardening at higher concentrations. This behaviour has been attributed to the competing effects of hydrogen, which enhances kink pair nucleation while hindering kink migration \cite{Katzarov2017-ei, Itakura2013-gt, Kim2025-bu, Huang2023-la,Leveau2025-ie, Matsui1979-om}. Similar effects have been observed for carbon impurities in iron using classical interatomic potentials \cite{Wang2019-nh, Allera2022-iv}. Understanding helium–dislocation interactions in this dilute regime provides a foundation for future investigations at higher helium concentrations that are representative of irradiated tungsten.

We find that both individual helium atoms and small helium clusters bind more strongly to dislocation kinks than to straight dislocation segments, with substantially stronger binding to vacancy kinks than to interstitial kinks. For a single helium atom, preferential stabilisation of the vacancy kink substantially reduces the kink pair nucleation barrier. On the other hand, a helium atom encountered by a migrating kink can act as a trap. In the case of $\mathrm{He}_\mathrm{2}$ and $\mathrm{He}_\mathrm{3}$ clusters, we identify a mechanism in which the kinked dislocation line is the ground state and the straight dislocation line is a metastable intermediate along the minimum energy path for dislocation motion. The energy barriers for this mechanism are lower than the kink pair nucleation in the pure metal. Molecular Dynamics (MD) simulations at 900~K show a softening effect in the case of $\mathrm{He}_\mathrm{1}$ and $\mathrm{He}_\mathrm{2}$ under the conditions studied. 

This is accomplished by building a \ac{MLIP} from \textit{ab initio} data for low concentration helium in tungsten. The large time and length scales that are required to study dislocation mobility are well beyond the limits of density functional theory (DFT), due to its computational cost which typically scales cubically with the number of electrons. Kink pairs break the symmetry along the dislocation line,  thus ruling out the possibility of a short simulation cell that is periodic along the dislocation line. Additionally, large cells are required to limit the self-interaction of impurities through periodic boundaries and allow for long-range relaxation of the dislocation core (i.e., the region immediately surrounding dislocation lines, where the crystal structure is most distorted).  

Recent work with hybrid quantum-mechanical/molecular-mechanical (QM/MM) methods has allowed for a quantum accurate description of impurity-dislocation interactions on a larger length scale \cite{Grigorev2023-fm}. However, these simulations require force integration to obtain energy differences and are restricted in time by the cost of the DFT calculation used in the quantum region and the need for large `buffer' regions to mitigate the effects of artificial surfaces \cite{Swinburne2017-oy}. Classical interatomic potentials may provide access to larger time and length scales, but generally they are not accurate enough and there are often significant discrepancies when simulation results are compared to those from DFT \cite{Marinica2013-mp,Bonny2014-xz, Allera2025-js}. Using \acp{MLIP} provides access to the larger time and length scales, whilst aiming to maintain quantum accuracy.

\section{Methodology}
\subsection{Building the \textit{ab initio} database and fitting the potential}
Using the linear Atomic Cluster Expansion (ACE) fitting code \texttt{ACEpotentials.jl (v0.6.5)} \cite{Witt2023-eu}, a potential for the W-He system was developed (with correlation order 3, a maximum polynomial degree of 21, and a cutoff of 5.0~\AA). This builds upon the training set used for an existing \ac{MLIP} for screw dislocations in pure tungsten \cite{Szlachta2014-un}, with the aim of improving the accuracy of dislocation kinks and expanding to the two-species system to study helium in bulk and helium-induced dislocation core reconstructions. This required the training database to include configurations with relevant atomic environments. 

To build dislocation configurations efficiently and keep the number of atoms low, we use a quadrupolar periodic array of dislocation dipoles \cite{Ventelon2007-if}. It was found that the widely used 135~atoms/$\it{b}$ simulation cell could be compressed further into 45~atoms/$\it{b}$ (built following the convention in \cite{Ventelon2013-zo} with $n,m = 9,5$). The dislocations are separated by approximately $12~\mathrm{\AA}$ in the 45~atoms/$\it{b}$ geometry (versus $19~\mathrm{\AA}$ for 135~atoms/$\it{b}$), which if used for direct \textit{ab initio} studies would result in undesirable finite-size effects \cite{Clouet2009-ox}. However, the minimally strained tungsten atomic environments that separate the dislocation cores in the 135~atoms/$\it{b}$ cell are adequately covered elsewhere in the database. The compressed cells have proven to be invaluable for collecting energy and force data for atomic environments near dislocation cores, for otherwise unobtainable dislocation line and kink sizes (up to 10$\it{b}$ in length). Testing confirmed that \acp{MLIP} trained on 45~atoms/$\it{b}$ configurations make predictions on unseen 135~atoms/$\it{b}$ configurations with reduced errors compared to those without them.

All DFT configurations in this work were evaluated using the \texttt{CASTEP} plane-wave DFT code \cite{Clark2005-bs}, using versions 20 and 22 which were verified to be consistent, and version 23, which had a negligible energy difference of approximately $10^{-5}$~eV/atom compared to the previous versions. Calculations employed the Perdew–Burke–Ernzerhof (PBE) exchange–correlation functional within the generalised gradient approximation \cite{Perdew1996-sy}. The DFT database from \cite{Szlachta2014-un} was reevaluated due to unresolvable energy differences, likely arising from changes in \texttt{CASTEP} over the past decade, and to allow for the use of a higher plane-wave cut-off energy of 700~eV (previously 600~eV), which is more appropriate for helium. This choice is conservative, as most relative energy differences --- and hence forces --- converge before this value. The parameter \texttt{fine\_grid\_scale}, which controls the density of the fine FFT grid relative to the coarse grid, was set to 2.
An ultrasoft pseudopotential was chosen for tungsten that treats 14 electrons ($5\mathrm{s}^2$ $5\mathrm{p}^6$ $5\mathrm{d}^4$ $6\mathrm{s}^2$) as valence after finding that 6 electrons ($5\mathrm{d}^4$ $6\mathrm{s}^2$) was not sufficient and 28 electrons ($4\mathrm{f}^{14}$ $5\mathrm{s}^2$ $5\mathrm{p}^6$ $5\mathrm{d}^4$ $6\mathrm{s}^2$) provided negligible benefit. A smearing width of 0.05~eV was used. A $k$-point spacing of $0.015~\text{\smash{\AA}}^{-1}$ was used for primitive cells, from which stresses are learned. The remainder of the data was computed at a $k$-point spacing of $0.025~\text{\smash{\AA}}^{-1}$ to reduce computational expense, particularly required since cells with hundreds of atoms were routinely studied. The convergence requirement for self-consistency on the total energy was $10^{-8}$~eV. The DFT training data and fitted ACE potential are available in the associated Zenodo repository.

\subsection{Atomistic modelling of screw dislocation mobility}
\begin{figure}
        \centering
        \includegraphics[width=\linewidth]{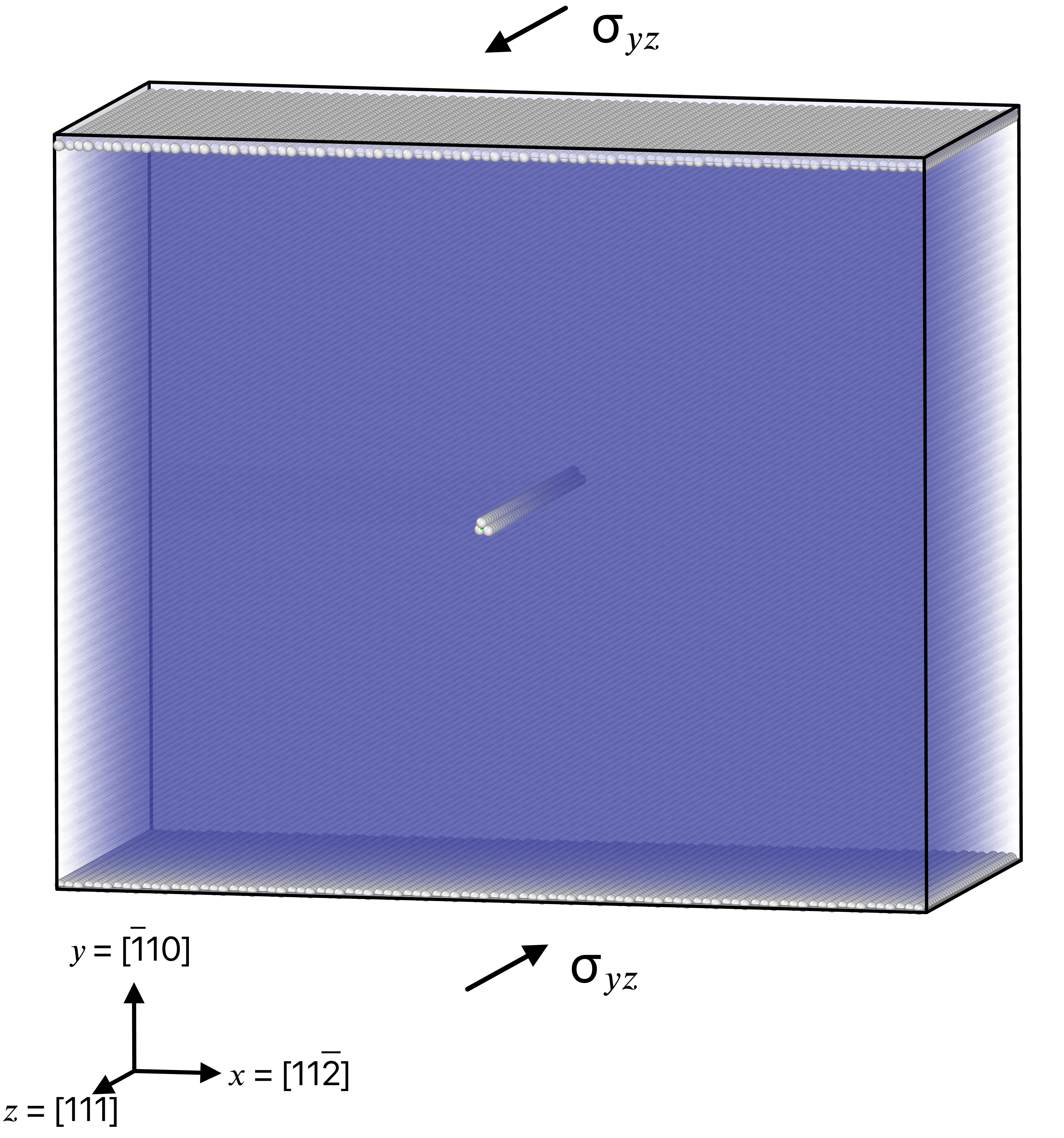}
        \caption{Simulation cell used to study screw dislocation motion on \{110\}. Dimensions in $x$, $y$ and $z$ are $235\,\mathrm{\AA} \times 200\,\mathrm{\AA} \times 111\,\mathrm{\AA}\ (40\,\textit{b}$), respectively, with periodic boundary conditions applied in $x$ and $z$. When shear stress is applied, it is done by force-loading the atoms in the top and bottom free surfaces that are coloured white. The dislocation core is also shown in white, while the bulk atoms are shown in transparent blue.}
        \label{fig:paperpadstress}    
    \end{figure}

Simulations of screw dislocation mobility on \{110\} planes --- both static and dynamic, with and without helium --- were performed using a periodic array of dislocations \cite{Bacon2009-wd}.  This is known to require approximately $10^5$ atoms due to various finite-size effects, particularly for MD simulations \cite{Cereceda2012-fd,Allera2025-js, Bacon2009-wd, Leveau2025-ie}. In this work, the dimensions of the simulation cell in the glide direction (along $ x  = [11\overline{2}]$), glide plane normal (along $ y  = [\overline{1}10]$), and dislocation line (along $ z  = [111]$) are chosen to be $235~\mathrm{\AA} \times 200~\mathrm{\AA} \times 111~\mathrm{\AA}\ (40~\textit{b}$), respectively, for a total of 316,800 atoms (see Figure \ref{fig:paperpadstress}). The dislocation is placed at the centre of the simulation cell, using displacements derived from the continuum theory of anisotropic elasticity. Periodicity is imposed in $x$ and $z$. There are free surfaces in $y$, where shear stress $\sigma_{yz}$ is applied through force-loading. Enforcement of periodic boundary conditions in the glide direction results in additional shear $\sigma_{xz}$ which can be minimised by increasing the size of the cell in the direction of the glide plane normal, but instead the cell was minimised with unfixed periodicity vectors. 

Energy (and enthalpy) barriers were calculated using the Nudged Elastic Band (NEB) method. Under applied stress, the relevant enthalpy is obtained by subtracting the work performed by the boundary-loading forces from the potential energy, $\Delta H = \Delta E - W$. Initial and final images were minimised to force tolerances of $10^{-4}$~eV/\AA. The NEB simulations had a force tolerance of $10^{-3}$~eV/\AA, a spring constant of $1.0~\text{eV/\smash\AA}^2 $, and a number of intermediate images that varied with mechanistic complexity. The nucleation and migration of kinks on the dislocation line requires $\gtrsim20$ images to achieve the desired force convergence. Simpler processes, such as the diffusion of an interstitial atom, can be studied with as few as 5 images, although 11 were used here for illustrative purposes. Dislocation core positions were calculated using the cost function method implemented in the \texttt{matscipy.dislocation} module \cite{Grigorev2024-zk}. In the NEB plots, the energy profiles are interpolated using a cubic Hermite spline, which passes exactly through the image energies, and the gradient along the reaction coordinate at each image was obtained by projecting the atomic forces onto the path tangent. Interaction energies were calculated using cylindrical cells with radius 40~\AA\ (with an additional 10~\AA\ fixed buffer region) and 50~\textit{b} length, minimised to a force tolerance of $10^{-4}$~eV/\AA. The atomic simulation environment (ASE) was used to handle atomic configurations and run these static simulations \cite{Hjorth_Larsen2017-wg}, with an emphasis on \texttt{ase.calculators.lammpslib} (which depends on \texttt{LAMMPS} \cite{Thompson2022-tm}), and the preconditioned minimisers in the \texttt{ase.optimize.precon} module \cite{Packwood2016-nj,Makri2019-dd}.

For MD calculations, using \texttt{LAMMPS} \cite{Thompson2022-tm}, the simulation cell was rescaled to account for the thermal expansion of the lattice parameter. A 1~fs timestep was chosen for simulations involving helium, and 2~fs for pure tungsten simulations. Shear stress $\sigma_{yz}$ was applied and the cell was equilibrated to the chosen temperature over 50~ps with a Langevin thermostat (3~ps damping parameter). Subsequently, the thermostat was removed and a 200 ps production run was performed in the NVE ensemble, from which dislocation positions were extracted (using OVITO dislocation analysis \cite{Stukowski2009-cv}) and an average dislocation velocity was calculated. The work done due to dislocation motion is negligible (in the fastest moving case, this leads to a $\sim1\%$ increase in temperature), so there is no need for a thermostat.

The computational cost of ACE potentials scales as $N^v$ with the number of chemical species $N$, with correlation order $v=3$ used here. This scaling can be reduced with compression \cite{Darby2023-gy}, although this was not available at the time of writing. Given that only a small section of our simulation cell contains any helium atoms, it would be wasteful to use the two-species potential over the entire simulation domain. The \texttt{LAMMPS} plugin \texttt{ML-MIX} \cite{Birks2025-bv} was used to model the environment around the helium cluster with the two-species W-He potential, and the remainder of the simulation cell with a one-species W potential (derived from the two-species potential). This was numerically identical to using the two-species potential across the entire simulation domain, but resulted in computational savings on the order of $10^6$ CPU hours. To use our potential with \texttt{LAMMPS}, a fork of the \texttt{ML-PACE} package~\cite{Lysogorskiy2021-zz} was required, as described in the \texttt{ACEpotentials.jl} documentation \cite{Witt2023-eu}.

\section{Results}
\subsection{Dislocation core reconstructions}

    Impurities are drawn to dislocations from the bulk due to the attraction of their respective stress fields. The presence of impurities leads to the reconstruction of the dislocation core and the corresponding interaction energy $E_{\text{int}}$ can be calculated as 
    \begin{equation}
        E_{\text{int}} = E_{\mathrm{dis+He}_n} - E_{\text{dis}} - (E_{\mathrm{bulk+He}_n} - E_{\text{bulk}}),
    \end{equation}
    where $E_{\text{dis}}$ and $E_{\text{bulk}}$ are the energies of a pure tungsten simulation cell with and without a dislocation, respectively. $E_{\mathrm{dis+He}_n}$ and $E_{\mathrm{bulk+He}_n}$ are the energies of the same simulation cells, but with $\mathrm{He}_n$ at the most stable site.

In pure bcc metals, the ground state of the screw dislocation lies at the centroid of three [111] atomic columns, often referred to as a Peierls valley. The dislocation core structure in this case is known as the \emph{easy core}, and the presence of impurities leads to spontaneous reconstruction away from this ground state. This has been thoroughly investigated in the literature, for various metals, interstitial impurities, and the concentration of the impurity along the dislocation line \cite{Borges2022-ag,Grigorev2023-fm,Zhao2019-lq, Luthi2017-vh, Ventelon2015-tx, Hachet2020-iy, Luthi2019-jr}.

\begin{figure}[h]
        \centering
        \includegraphics[width=\linewidth]{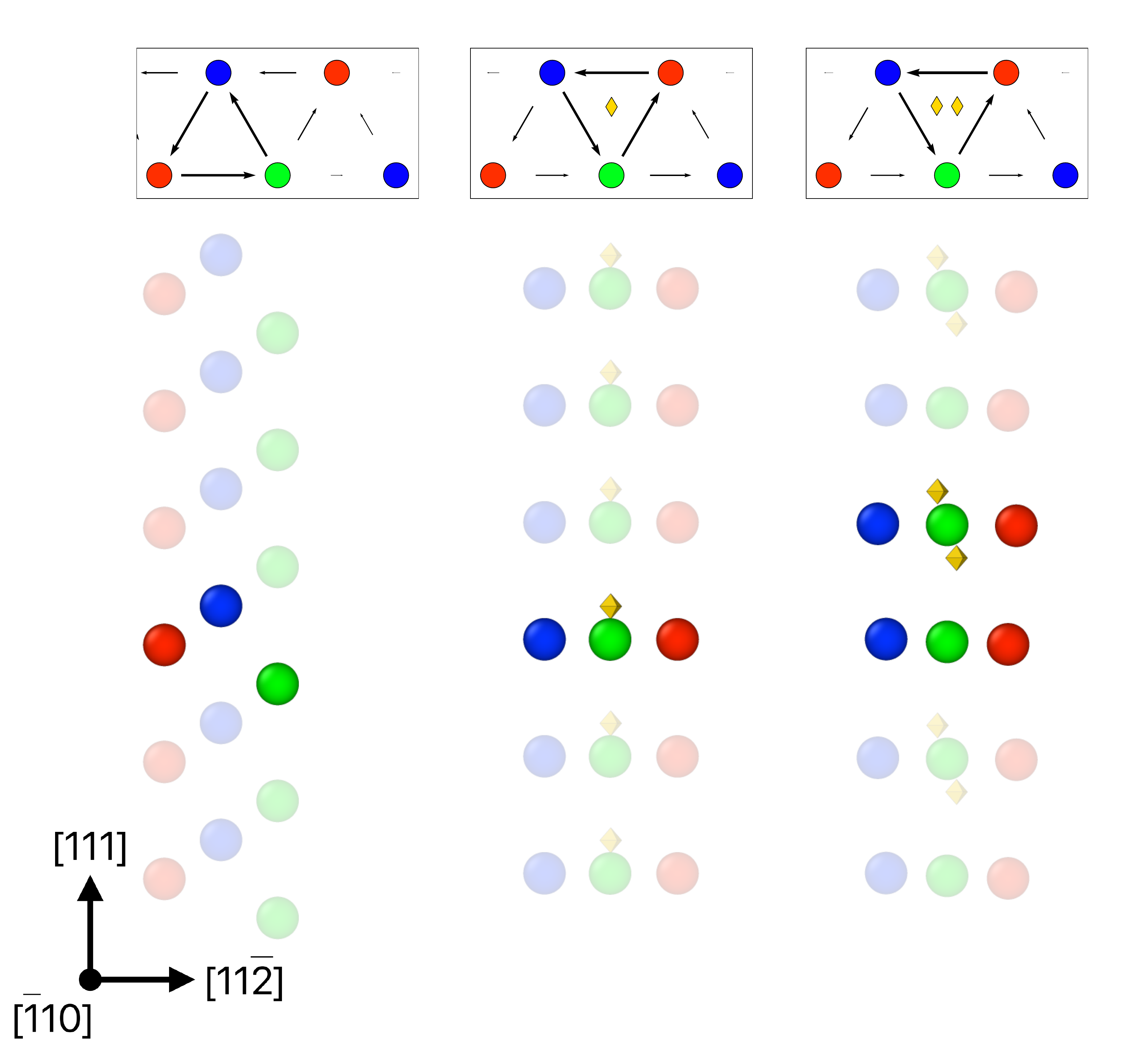}
        \caption{Reconstruction of the dislocation core in the presence of helium (saturated regime). From left to right: No He, 1He/$\it{b}$ and 2He/2$\it{b}$. Differential displacement maps (top) view down the dislocation and illustrate displacements of the columns relative to bulk.  W and He atoms are represented by spheres and diamonds, respectively. Clustered reconstruction (2He/2$\it{b}$) is $0.13$~eV lower in energy per He atom than the original periodic array (1He/$\it{b}$). Shaded out atoms correspond to periodic images.}
        \label{fig:corestructures}
    \end{figure}

    When the screw dislocation line is saturated with one helium atom per $\it{b}$, the dislocation reconstructs to the \emph{hard core}, where the atoms in the three atomic columns surrounding the core lie in the same plane \cite{Grigorev2023-fm}, as shown in Figure~\ref{fig:corestructures}. The hard core is a maximum in the two-dimensional potential energy landscape of the straight dislocation moving in the (111) plane of the pure metal \cite{Ventelon2013-zo}. If modelled in a 1$\it{b}$ long simulation cell, the helium atoms are constrained to exist in a periodic array, and the interaction energy of the helium atom, relative to a tetrahedral site in bulk, is calculated with our ACE potential to be $-1.46$~eV. 
    
    When the translational symmetry along the dislocation line is allowed to break (cell length $>$ 1$\it{b}$), we find that this periodic arrangement is not the most stable. In this relatively saturated regime, we find that it is preferential for helium atoms to form clusters (see right hand side of Figure \ref{fig:corestructures}), as is the case in bulk tungsten \cite{Becquart2006-wk}. In this case, in a 2$\it{b}$ long simulation cell, the interaction energy per helium atom is $-1.59$~eV, making this configuration 0.13~eV/He lower in energy than the $1b$ arrangement. This observation emphasises the need for large simulation cells (and therefore cheaper methods, such as MLIPs), even in relatively saturated regimes, due to the intricate periodicity-breaking arrangements that may form around the core.

     \begin{figure}
    \includegraphics[width=0.95\linewidth]{figures/jul2025allkinks2_alltogether.pdf}
    \caption{Structures of $\mathrm{He}_n$ bound to straight dislocations and kinks. Only atoms around the dislocation are shown, with highlighting to emphasise the character of the kinks. Dislocation lines (in green) are from OVITO dislocation analysis \cite{Stukowski2009-cv}. Insets show the view down the [111] direction.}
    \label{fig:kinkstructures}
\end{figure}

    Our simulations predict that in the dilute regime, where a single helium atom is situated within a sufficiently long dislocation line, the dislocation core reconstructs towards the \emph{split core} in the vicinity of the helium. This is another maximum in the two-dimensional energy landscape of the dislocation in the (111) plane of the pure metal, where the dislocation lies close to a $[111]$ atomic column. Away from the helium atom, the dislocation core structure gradually reverts to the easy type over a distance of approximately 10$\it{b}$, in agreement with previous QM/MM studies \cite{Grigorev2023-fm}.
    
    We calculate an interaction energy of $-1.70$~eV for the helium atom with the straight dislocation in the dilute regime, using our ACE potential. This differs slightly from the published QM/MM result of $-1.5$~eV \cite{Grigorev2023-fm}. Our DFT calculations yield a value of $-1.52$~eV with finite-size effects --- both in-plane and along the dislocation line --- that are expected to weaken the apparent binding. Therefore, we argue that the difference here, at least in part, is due to surface effects from the QM cluster, despite a buffer region, in the QM/MM case. 
        
    Using our ACE potential, the interaction energies of $\mathrm{He_2}$ and $\mathrm{He_3}$ with the straight dislocation line are calculated as $-$3.05~eV and $-$4.28~eV, respectively (structures shown in Figure~\ref{fig:kinkstructures}). The incremental interaction energy of bringing together helium atoms, which are already present on the dislocation line, is also considered and can be calculated as
    \begin{equation}
        E_{\text{inc-int}} = E_{\mathrm{dis+He_n}} + E_{\text{dis}} - E_{\mathrm{dis+He_{n-1}}} - E_{\mathrm{dis+He_{1}}},
    \end{equation}
    where $E_{\text{dis}}$ is the energy of a pure tungsten simulation cell containing a dislocation. $E_{\mathrm{dis+He_1}}$, $E_{\mathrm{dis+He_{n-1}}}$ and $E_{\mathrm{dis+He_{n}}}$ are the energies of the same simulation cells with a He cluster of size $1$, $n-1$, and $n$, respectively, located at the most stable site on the dislocation. 
    
    When two isolated helium atoms on the dislocation line come together to form a cluster, the incremental interaction energy is $-0.57$~eV, indicating that clustering is energetically favourable. The addition of a third isolated helium atom present on the line to the $\mathrm{He}_2$ cluster has an incremental interaction energy of $-0.82$~eV, demonstrating a further preference for clustering. These results slightly differ from those of the QM/MM study, which claims a minor repulsive interaction between helium atoms on the line \cite{Grigorev2023-fm}. However, it appears that the QM/MM study did not find the ground state of the helium cluster on the straight dislocation line. Given the significantly increased cost of the QM/MM simulations compared to using an interatomic potential, it will not have been feasible to explore as many starting configurations as we did with our potential.

     There are other minima with similar energies that exist for the configurations where the $\mathrm{He_2}$ and $\mathrm{He_3}$ clusters are bound to the straight dislocation, which will not be discussed. This is because they are all metastable, with the reconstruction of the dislocation line to the kink pair configuration being the most stable, due to the preference for the cluster to bind to dislocation kinks, which is discussed in the following sections.

\subsection{Interaction of helium with dislocation kinks}

Periodic arrays of individual dislocation kinks can be constructed using appropriately crafted simulation cells \cite{Ventelon2009-rq}, allowing the energy of their interaction with helium to be calculated. For $\mathrm{He_1}$, $\mathrm{He_2}$ and $\mathrm{He_3}$, we observe significant increases in binding compared to the straight dislocation for both types of kinks: interstitial and vacancy, the structures of which are shown in Figure \ref{fig:kinkstructures}. A noticeable asymmetry is also evident, with the vacancy kink binding helium substantially more strongly than the interstitial kink.

\begin{figure}
\centering
\begin{minipage}{\linewidth}
\centering
\renewcommand{\arraystretch}{1.2} 
\small
\begin{tabular}{|l|c|c|c|}
\cline{2-4}
\multicolumn{1}{c|}{} &
\multicolumn{3}{c|}{\textbf{\boldsymbol{$E_{\mathrm{int}}$} (\textbf{\boldsymbol{$E_{\mathrm{pref}}$}}) / eV}} \\ 
\cline{2-4}
\multicolumn{1}{c|}{} &
\textbf{Straight disloc} &
\textbf{Interstitial kink} &
\textbf{Vacancy kink} \\ \hline
\boldsymbol{$\mathrm{He}_1$} & $-1.70$ & $-1.99$ ($-0.29$) & $-2.81$ ($-1.11$) \\ \hline
\boldsymbol{$\mathrm{He}_2$} & $-3.05$ & $-3.60$ ($-0.55$) & $-5.03$ ($-1.98$) \\ \hline
\boldsymbol{$\mathrm{He}_3$} & $-4.28$ & $-5.20$ ($-0.92$) & $-6.67$ ($-2.39$) \\ \hline
\end{tabular}

\vspace{0.2cm}

\justifying
\setlength{\parindent}{0pt}

\small\textbf{(a)} Interaction energies calculated with ACE in $\approx10^5$ atom cylindrical cells. 
Values in parentheses denote the preference to bind to kinks relative to the straight dislocation.
\end{minipage}

\vspace{0.4cm}

\includegraphics[width=\linewidth]{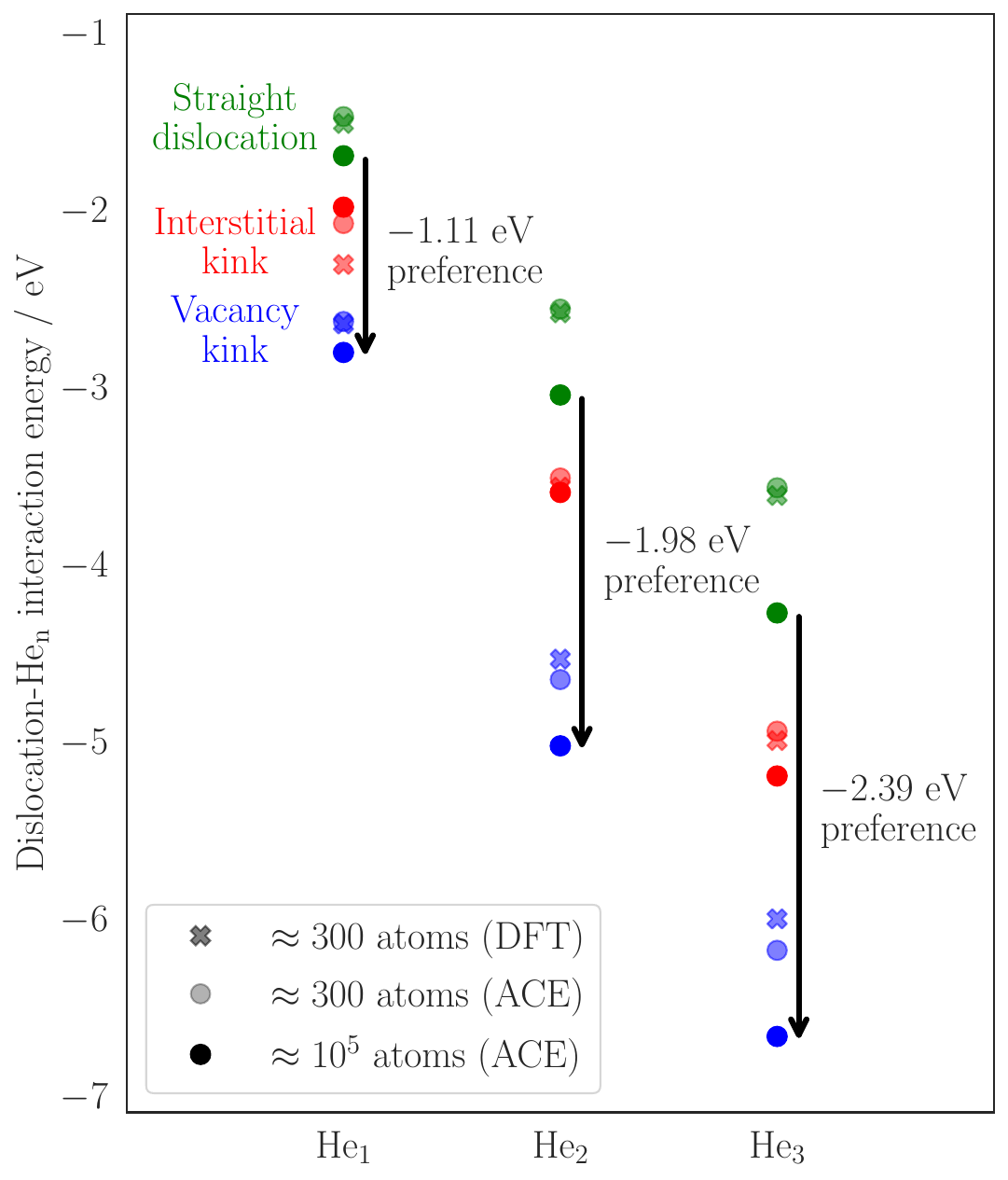}

\vspace{0.2cm}

\begin{minipage}{\linewidth}
\justifying
\small
\setlength{\parindent}{0pt}

\textbf{(b)} Comparison of interaction energies calculated with ACE and DFT. 
DFT reference calculations (crosses) use 45~atoms/$b$ quadrupole cells, with lengths of $6b$ for straight dislocation and vacancy kink configurations and $8b$ for interstitial kink configurations; the same cells are also evaluated with ACE (light circles). Larger ACE calculations with $\approx 10^5$ atoms in cylindrical cells to remove finite-size effects are also shown (dark circles). 
\end{minipage}

\caption{Interaction energies of $\mathrm{He}_n$ clusters with screw dislocations and kinks, relative to helium in the most stable site in bulk tungsten. Bulk references use 432(+$n$) atom cells.
(a) Calculated with ACE in $\approx 10^5$ atom cells. 
(b) DFT versus ACE comparison of the interaction energies in $\approx 300$ atom cells, and the ACE value from the $\approx 10^5$ atom cells.}
\label{fig:He_kink_interaction}
\end{figure}

For the single helium atom, the interaction energies with the interstitial and vacancy kinks are $-1.99$~eV and $-2.81$~eV, respectively. Therefore, the preference for kink binding over the straight dislocation is $E_{\text{pref,IK}}\approx-0.3$~eV and $E_{\text{pref,VK}}\approx-1.1$~eV for the interstitial and the vacancy kink, respectively. The magnitude of the kink binding preference increases with cluster size (Figure \ref{fig:He_kink_interaction}a).
For $\mathrm{He_2}$ and $\mathrm{He_3}$ the vacancy kink preferences are 1.98~eV and 2.39~eV in magnitude, respectively, exceeding the 1.58~eV kink pair nucleation energy and making the kink pair configuration the ground state.

These large interaction energies allow us to partially validate our results in small cells solely with DFT (Figure \ref{fig:He_kink_interaction}b), since the finite-size errors do not mask the qualitative behaviour. Values are obtained using 45~atoms/$b$ quadrupolar cells with dislocation line length $\ge6 b$, calculated both with DFT and ACE. Good agreement is observed in all cases, although the largest discrepancies occur for the helium-bound kink structures. Our ACE potential predicts weaker binding for the single helium atom with the interstitial kink and stronger binding for $\mathrm{He_2}$ and $\mathrm{He_3}$ with the vacancy kink, relative to DFT.

The greater affinity of the vacancy kink for helium impurities is likely attributed to the free volume being concentrated in one place, as opposed to the interstitial kink where there is free volume off-centre on both sides (although this is not quantified). This can be better understood by looking at the structural changes of the three $[111]$ atomic columns around the kink (highlighted in Figure \ref{fig:kinkstructures}). We also observe that a tungsten vacancy point defect preferentially associates with the locally compressive region of the kinks, opposite to the preference of interstitial helium for regions of free volume, but for quantitative analysis the training database would need to be further extended with relevant representative configurations.

\begin{figure*}
    \centering
    \includegraphics[width=\textwidth]{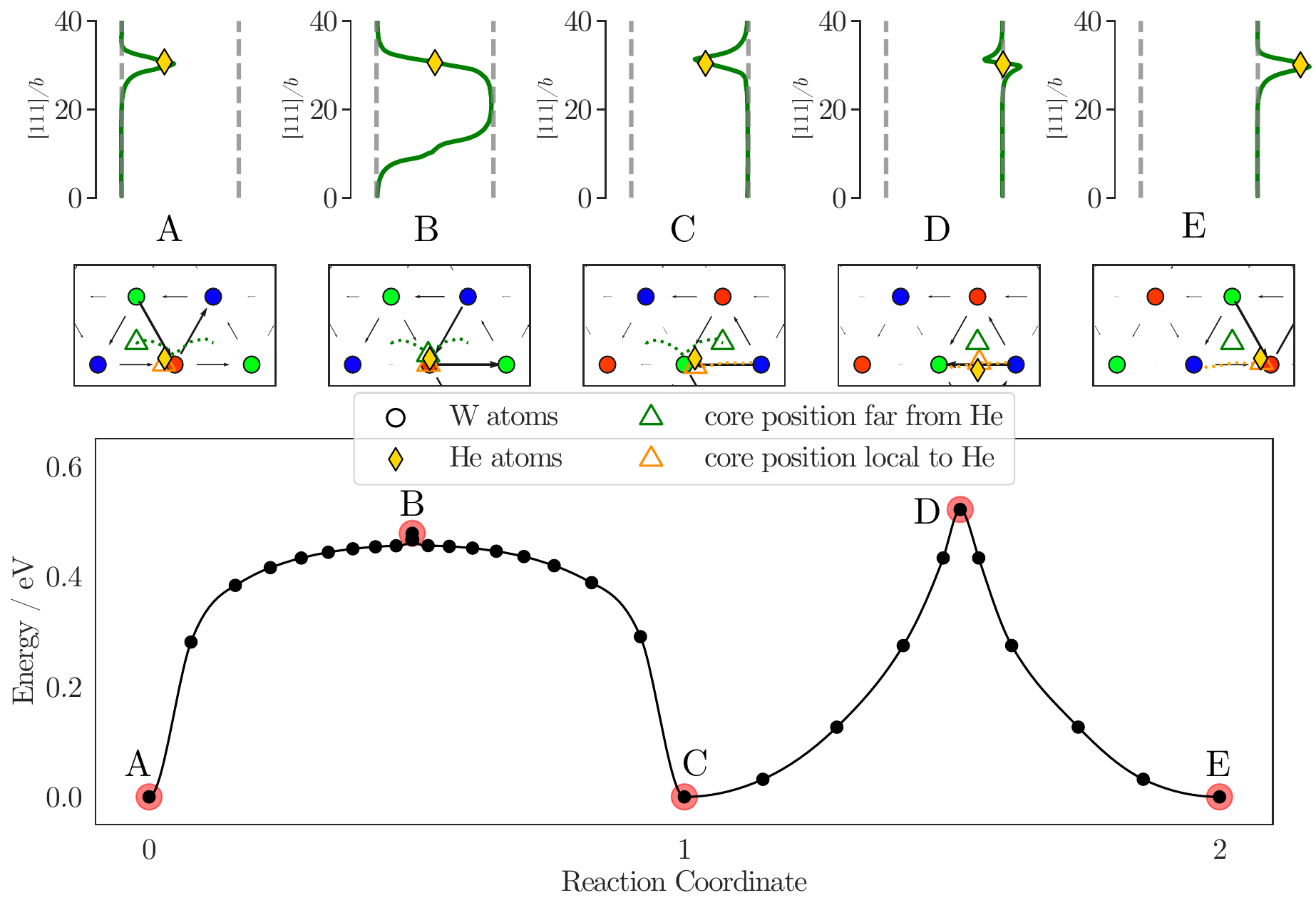}
    \caption{Two-step NEB pathway for helium-assisted kink pair nucleation. A$\rightarrow$C: Kink pair nucleation through helium-stabilised vacancy kink, with interstitial kink migration. C$\rightarrow$E: Diffusion of helium to the next [111] atomic column. Dislocation line positions are determined using the cost function method, as they move between Peierls valleys (grey dashed lines), and differential displacement maps (for the layer that is local to helium) are shown for the key images. In the Vitek maps, the colour of the [111] atomic columns changes across the image series as the helium atom moves between layers. Green and orange dashed lines on the differential displacement maps represent the path taken by the dislocation far (20~\textit{b}) from and local to the helium, respectively. The triangles correspond to the position in the current image. The reaction coordinate is normalised by average dislocation position for A$\rightarrow$C and C$\rightarrow$E.}
    \label{fig:NEB1}
\end{figure*}
    \subsection{Movement of helium on the dislocation line}
    A single helium atom will diffuse along the dislocation line in a helical manner, moving between the minima next to each of the three [111] atomic columns that surround the dislocation line (see Figure~\ref{fig:hediffhelix_combined}). For each hop, the position of the helium atom along the dislocation line changes by $b/3$ and after three hops the atom will be in the same position in the (111) plane as when it started, but will have advanced by $b$ along [111]. For comparison, both the present ACE and DFT calculations give a helium interstitial migration barrier in bulk tungsten of approximately 0.06~eV, consistent with the DFT value reported in ~\cite{Becquart2006-wk}. The corresponding helical-diffusion barrier along the dislocation line is substantially larger, with the ACE potential giving a value of 0.52~eV. Diffusion along the straight dislocation line is considered only for a single helium interstitial because the straight dislocation configurations containing $\mathrm{He}_n$ clusters with $n>1$ are metastable with respect to the kink pair configurations.
       
    There are also diffusion events at dislocation kinks, to move a helium atom between the equivalent sites that exist on both sides of the kinks. Since the position of the helium at the vacancy kink is close to the central atomic column, its transition from one side to the other involves only a very small displacement, with a barrier of $22$~meV, as shown in Figure~\ref{fig:hediffrk} and calculated with the ACE potential. The presence of this small barrier is believed to be representative of the underlying DFT, as DFT geometry optimisations show that the ground state of the helium atom at the vacancy kink site is not located at the exact centre of the free volume. As for the interstitial kink, where the helium is not positioned at this central atomic column, the diffusion event involves moving the helium atom by a distance equal to the distance between neighbouring Peierls valleys. Our ACE potential finds that this process has an energy barrier of 0.62~eV, as shown in Figure~\ref{fig:hedifflk}.

\subsection{Dislocation mobility in the presence of a single helium atom}
\begin{figure*}
    \centering
    \includegraphics[width=1\linewidth]{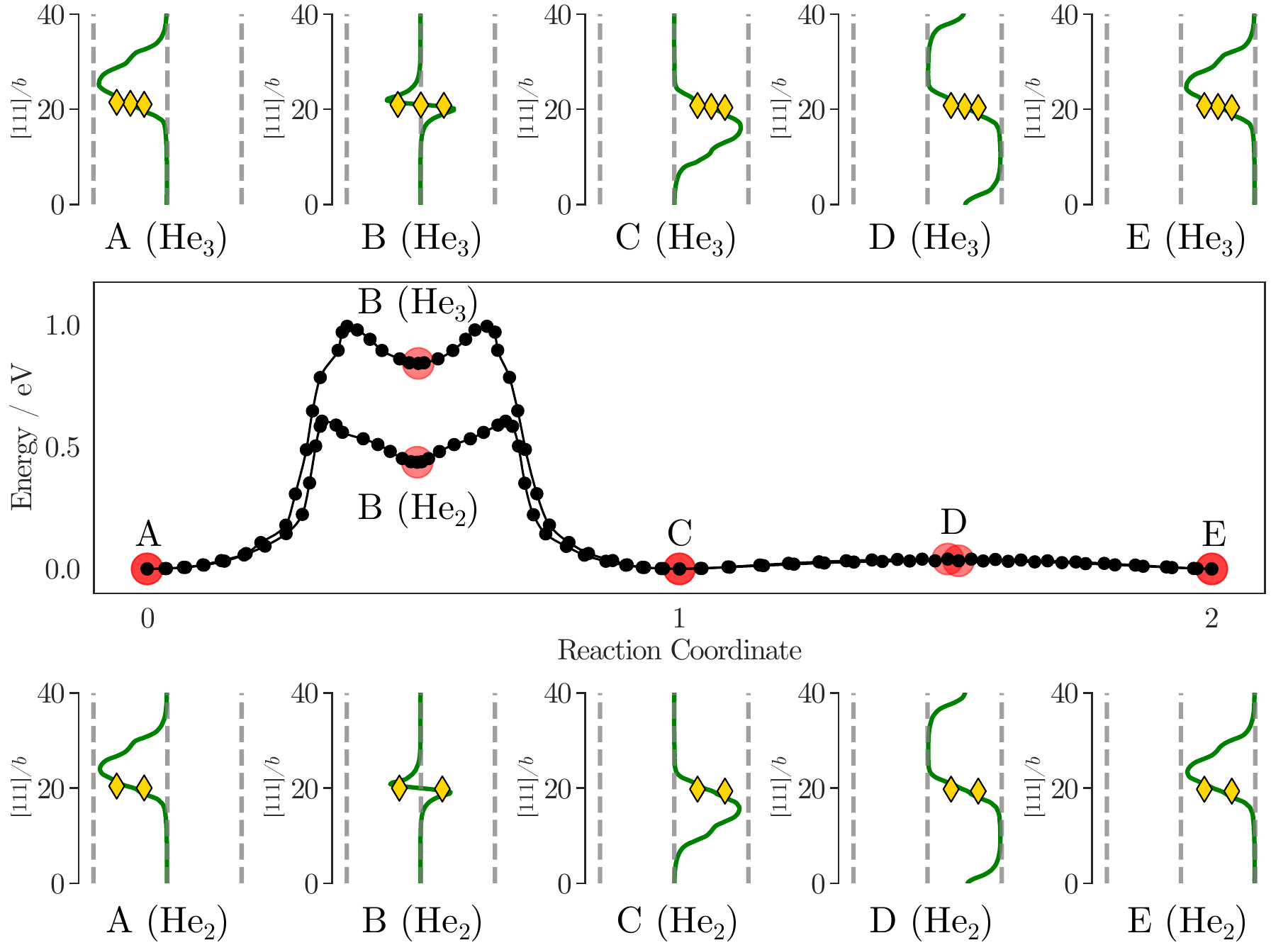}
    \caption{Energy profile from NEB pathway for dislocation motion in the presence of the $\mathrm{He}_2$ (bottom) and $\mathrm{He}_3$ (top). A$\rightarrow$B: Coupled migration of the helium cluster and reconstruction of the kink pair ground state to the metastable straight dislocation configuration. B$\rightarrow$C: the symmetry-related counterpart of A$\rightarrow$B, reconstructing the straight dislocation into an equivalent kink pair minimum. C$\rightarrow$E: Interstitial kink diffusion. Dislocation line positions (determined using the cost function method \cite{Ventelon2013-zo}) as they move between Peierls valleys (grey dashed lines) are shown for the key images. The reaction coordinate is normalised by average dislocation position for A$\rightarrow$C and C$\rightarrow$E.}
    \label{fig:he23mech}
\end{figure*}

In the case of an isolated helium atom, the favourability of binding to the dislocation kinks leads to a proportionate decrease in the kink pair nucleation energy, provided the mechanism proceeds with helium bound to one of the kinks. Using the NEB method \cite{Henkelman2000-yl}, we find that the most favourable mechanism involves helium binding to the vacancy kink, as expected due to stronger binding. For this to be observed, the helium must be situated next to the $[111]$ atomic column that separates the initial Peierls valley from the next. The reconstruction to the split core local to the helium atom means that the dislocation line is already partially moved towards the next valley. As shown in Figure \ref{fig:NEB1} (A$\rightarrow$C), the kink pair nucleates at the helium, forming a helium-stabilised vacancy kink and an interstitial kink that is free to move laterally to propagate the dislocation to the next Peierls valley. 

The kink pair nucleation energy in pure tungsten in this same simulation setup is 1.58~eV (in good agreement with another MLIP \cite{Goryaeva2021-de} and DFT-informed line-tension models \cite{Dezerald2015-ok, Clouet2021-pm}), which is reduced to 0.48~eV in the presence of the helium atom, differing by approximately $E_{\text{pref,VK}}$. There is a 22~meV perturbation in the middle of the path that corresponds to helium crossing the central $[111]$ column, since it does not sit exactly at the centre of the kink (see Figure \ref{fig:hediffrk}). To the best of our knowledge, this is the strongest reported example of impurity-assisted kink pair nucleation in a bcc metal. An approximate 10\% decrease in the nucleation energy has been observed in the Fe-H system, both with an embedded atom model \cite{Huang2023-la} and a neural network potential \cite{Meng2021-no}.

For this helium atom to repeatedly assist with dislocation motion, it must reposition itself adjacent to the next central [111] atomic column before each nucleation event (C$\rightarrow$E, Figure \ref{fig:NEB1}). This is the same diffusion event shown in Figure~\ref{fig:hediffhelix_combined}, and is one third of the helical cycle. During this diffusion event, the dislocation core local to the helium atom moves between split cores, whilst the dislocation as a whole remains in the same Peierls valley. The overall requirement for dislocation motion reduces from 1.58~eV in pure tungsten, to a pair of approximately 0.5~eV barriers in the presence of a single helium atom.

The mechanism that proceeds through the helium-stabilised interstitial kink can also be probed by carefully seeding the initial NEB image setup. However, the nucleation energy is only assisted by 0.29~eV (see Figure \ref{fig:NEB1lk}), due to the lower preference for interstitial kink binding. The kink pair nucleation energy in this case is reduced to 1.29~eV, differing from the pure tungsten barrier by approximately $ E_{\text{pref,IK}}$. The diffusion event to advance the helium atom to the next atomic column is different from the mechanism involving the pinned vacancy kink. We believe that the MEP involves the helium atom diffusing into the core after the dislocation has crossed the valley and the kink pair has annihilated, where the barrier to diffusion is only $0.1$~eV. Nevertheless, this is not believed to be a relevant pathway in this infinitely dilute regime due to the larger nucleation energy.

\subsection{Dislocation mobility in the presence of $\text{He}_2$ and $\text{He}_3$ clusters}

The clustering of helium atoms at the dislocation kinks further stabilises the kink relative to the straight dislocation (see Figure \ref{fig:He_kink_interaction}). The preference to bind to the vacancy kink is $-$1.98~eV and $-$2.39~eV for $\mathrm{He}_2$ and $\mathrm{He}_3$, respectively. Their magnitudes exceed the energy of kink pair nucleation ($1.58~\mathrm{eV}$), making the kink pair configuration energetically favourable. The helium-stabilised kink pair configurations lie 0.44 and 0.84~eV below the corresponding straight-dislocation configurations with the ACE model. The larger ACE--DFT discrepancies for helium-bound kink structures make the $\mathrm{He}_2$ ground state ordering less robust than that of $\mathrm{He}_3$.

The minimum energy path for dislocation motion in the presence of $\mathrm{He}_{2}$ and $\mathrm{He}_{3}$ goes between kink pair minima, via the metastable straight dislocation line, in concert with the movement of the helium cluster, as shown in Figure~\ref{fig:he23mech} (A$\rightarrow$C). The separation of the kink pair in the ground state is not equal to half the cell length, as is the case when a kink pair transition state is studied, because the favourable kink-kink interactions lower the cell energy. The kinks get as close as possible before the kink pair (and hence the accommodating kink environment) collapses. 

As shown in Figure~\ref{fig:he23mech}, the barrier for the coupled migration of the helium cluster and reconstruction of the kink pair ground state to the metastable straight dislocation is 0.60~eV for $\mathrm{He}_{2}$ and 1.00~eV for
$\mathrm{He}_{3}$. These barriers are considerably smaller than the kink pair nucleation energy in the pure metal. The second event in Figure~\ref{fig:he23mech} (C$\rightarrow$E) is migration of the free-to-move interstitial kink, required for the whole dislocation to have advanced by an entire Peierls valley. This barrier is small (approximately $40~\mathrm{meV}$), as it corresponds to the relatively weak attraction of opposite kinks, and will easily be overcome by a minor applied shear stress. It is calculated to be the same for both $\mathrm{He}_{2}$ and $\mathrm{He}_{3}$, due to the similar dislocation line configuration in both cases.

\subsection{Enthalpy barriers for dislocation mobility}

\begin{figure}
    \centering
    \includegraphics[width=1\linewidth]{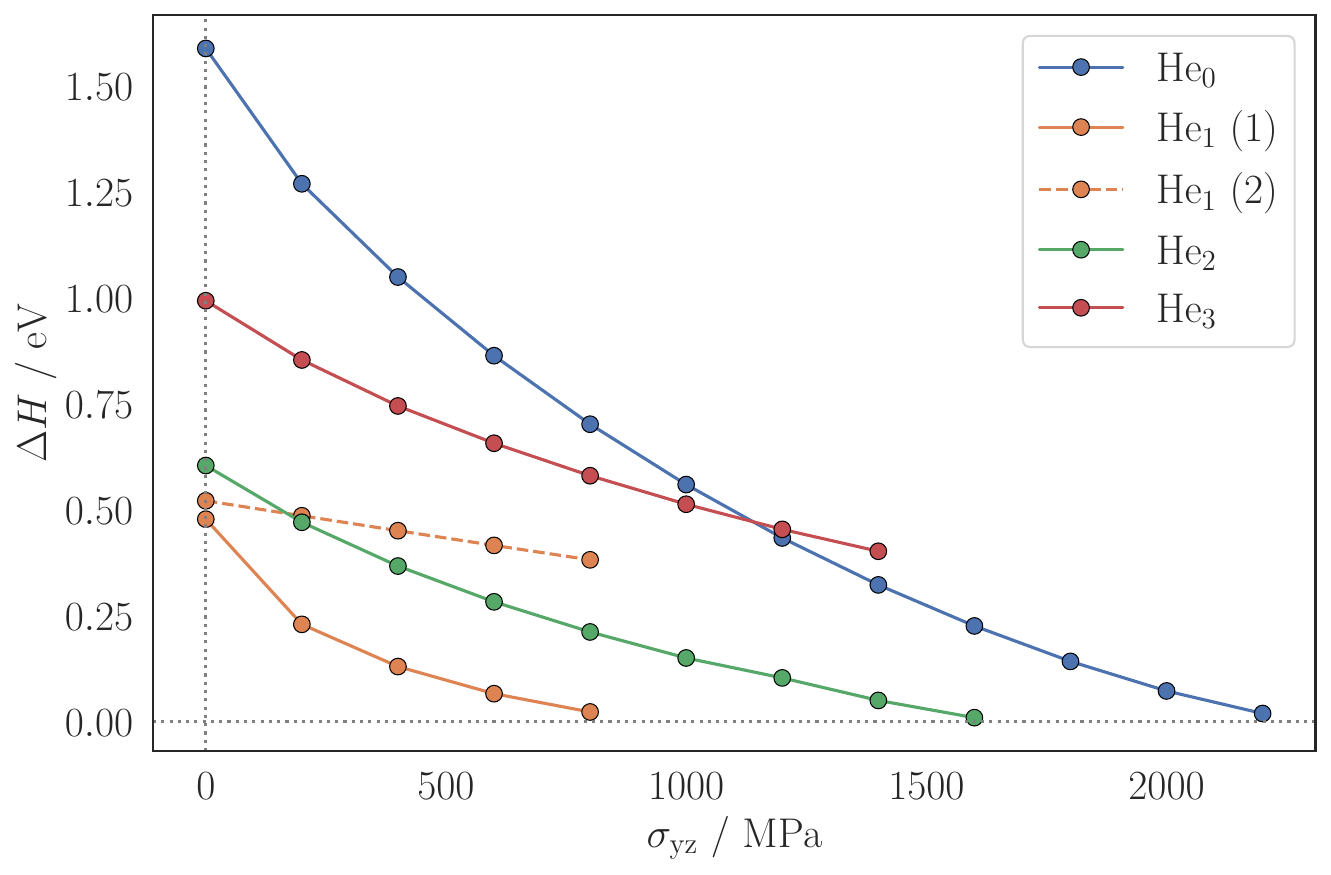}
    \caption{Enthalpy barriers for rate determining processes as a function of applied shear stress $\sigma_{yz}$ for dislocation motion in the presence of $\mathrm{He}_n$ for $0\le n\le3$.}
    \label{fig:stressrel}
\end{figure}

Figure~\ref{fig:stressrel} shows the stress dependence of the principal enthalpy barriers considered in this work: kink pair nucleation in pure tungsten, denoted $\mathrm{He}_{0}$; helium-assisted kink pair nucleation and subsequent intercore diffusion for $\mathrm{He}_{1}$, denoted $\mathrm{He}_{1}(1)$ and $\mathrm{He}_{1}(2)$, respectively; and transitions between kink pair minima for $\mathrm{He}_{2}$ and $\mathrm{He}_{3}$. The subsequent interstitial kink migration events in the $\mathrm{He}_{2}$ and $\mathrm{He}_{3}$ mechanisms are not shown due to their negligible energy barrier.

The enthalpy barrier for kink pair nucleation in pure tungsten is more sensitive to the applied shear stress than the main dislocation motion processes involving helium. It should be noted that the intercore diffusion of $\mathrm{He}_{1}$ varies only weakly with the applied shear stress, since only a small section of the dislocation line advances during the process. 

Up to approximately 400~MPa, the straight dislocation configuration remains a local minimum along the $\mathrm{He}_{2}$ and $\mathrm{He}_{3}$ pathways (image B in Figure~\ref{fig:he23mech}). Above this stress, the applied shear stress tilts the enthalpy profile sufficiently that the configuration loses metastability and relaxes directly towards image C. The small C$\rightarrow$E interstitial kink migration barrier is also eliminated, so image C no longer remains metastable and the configuration relaxes directly to image E. Barriers at higher stresses were therefore obtained using a direct A$\rightarrow$E NEB calculation, rather than by joining separate A$\rightarrow$B and B$\rightarrow$C paths through the straight dislocation minimum. For $\mathrm{He}_{3}$, the conventional pure tungsten mechanism has a lower enthalpy barrier at approximately 1000~MPa. Beyond this stress, a much stricter starting set of intermediate images is required if the helium-controlled mechanism is to be observed, since it is no longer the minimum enthalpy path; these data points are omitted from Figure~\ref{fig:stressrel}.

Taken together, the lower enthalpy barriers of the helium-assisted mechanisms relative to conventional kink pair nucleation suggest softening in the low-stress, single-cluster limit.

\subsection{Molecular dynamics simulations of dislocation mobility}
\begin{figure}
    \centering
    \includegraphics[width=1\linewidth]{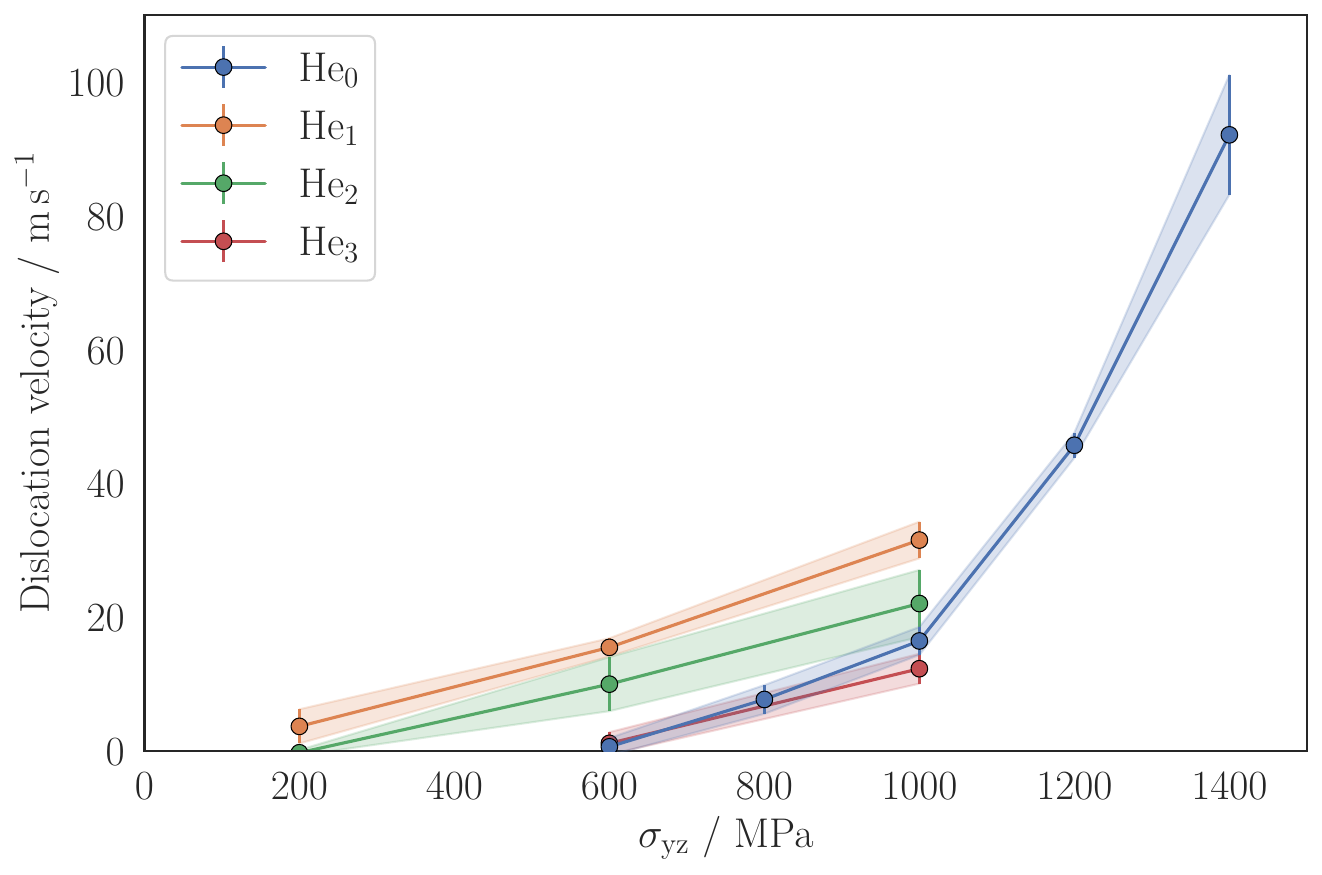}
    \caption{Dislocation velocity as a function of applied shear stress $\sigma_{yz}$ in the presence of $\mathrm{He}_n$ for $0\le n\le3$, from 900~K MD simulations. Results are the mean of three independent runs, with error bars representing $\pm2\times$ standard error.}
    \label{fig:dislocvels}
\end{figure}

Enthalpy barriers do not by themselves determine finite temperature rates, which also depend on entropic contributions and kinetic prefactors \cite{Allera2025-js}. Therefore, we performed MD simulations at 900~K to test whether the trends in Figure~\ref{fig:stressrel} translate into enhanced dislocation mobility. As shown in Figure \ref{fig:dislocvels}, dislocation motion is observed at substantially lower $\sigma_{yz}$ in the presence of $\mathrm{He}_{1}$ or $\mathrm{He}_{2}$ than in pure tungsten, demonstrating enhanced screw dislocation mobility. Analysis of the MD trajectories showed that this enhancement occurred through the mechanisms identified by the NEB calculations.

No comparable mobility enhancement is observed for $\mathrm{He}_{3}$ under the conditions studied. When $\sigma_{yz}=1000$~MPa, the $\mathrm{He}_{0}$ and $\mathrm{He}_{3}$ barriers are similar and conventional kink pair nucleation can occur at many more positions along the dislocation line than the localised $\mathrm{He}_{3}$-assisted event. The conventional mechanism therefore out-competes the helium-assisted pathway, after which the $\mathrm{He}_{3}$ cluster is dragged along with the moving dislocation and hinders its overall motion. If dislocation motion at lower applied stresses was computationally feasible then it is plausible that $\mathrm{He}_{3}$ would still have a softening effect. This could be investigated by running the MD at a higher temperature, but would require extension of the MLIP database to include relevant configurations.

Above 1000~MPa, the $\mathrm{He}_{0}$ mechanism also competes with the helium-assisted mechanisms $\mathrm{He}_{1}$ and $\mathrm{He}_{2}$. Additionally, in this high-stress regime, the time required for kink migration is no longer negligible relative to the kink pair nucleation time. Retention of the helium cluster at the vacancy kink therefore impedes the advance of the kink pair and reduces the overall dislocation velocity. At sufficiently high stresses, the dislocation can unpin from the helium impurity, after which its velocity approaches that of the dislocation in pure tungsten. 

During some unpinning events, helium clusters initially bound to vacancy kinks retained a vacancy-like configuration as the dislocation moved away, producing a helium-vacancy complex and a self-interstitial $[111]$ crowdion. The helium-vacancy complex remained immobile over the remainder of the 200~ps simulation, presumably due to its high migration barrier. However, the self-interstitial crowdion is highly mobile, as is known in the literature \cite{Wang2021-tx}. Although the MLIP training database contains some self-interstitial configurations, it is unlikely to cover the atomic environments associated with this defect-emission process. For these reasons, data points above 1000~MPa involving helium are not included in Figure \ref{fig:dislocvels}. 

Similarly, above 1400~MPa there is a significant increase in cross kinking, which leads to debris formation, which the dislocation unphysically encounters due to periodicity in the glide plane; the corresponding $\mathrm{He}_{0}$ data points are also excluded. A more thorough investigation of these high-stress behaviours is left for further study.

Overall, our finite temperature simulations indicate a softening effect in the single-cluster limit at the lowest stresses accessible to the MD simulations. This effect becomes less pronounced with increasing cluster size and is likely to be transient as helium accumulates and larger clusters form. 

\subsection{Kink trapping beyond the single-cluster limit}
\begin{figure}
    \centering
    \begin{minipage}{\linewidth}
        \centering
        \includegraphics[width=\linewidth]{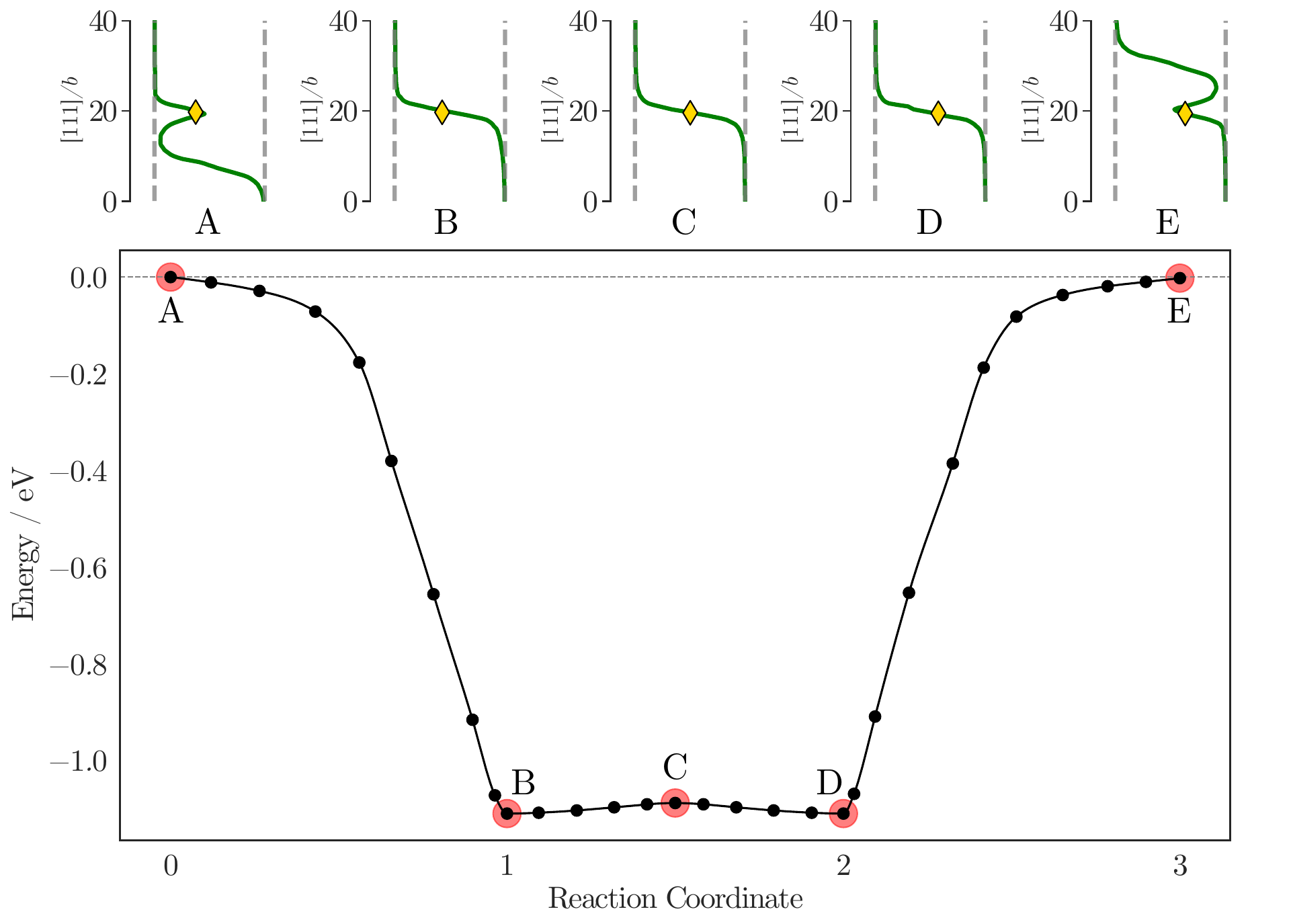}
        
        \textbf{(a)} Vacancy kink gliding past a helium interstitial.
    \end{minipage}

    \vspace{0.2cm}

    \begin{minipage}{\linewidth}
        \centering
        \includegraphics[width=\linewidth]{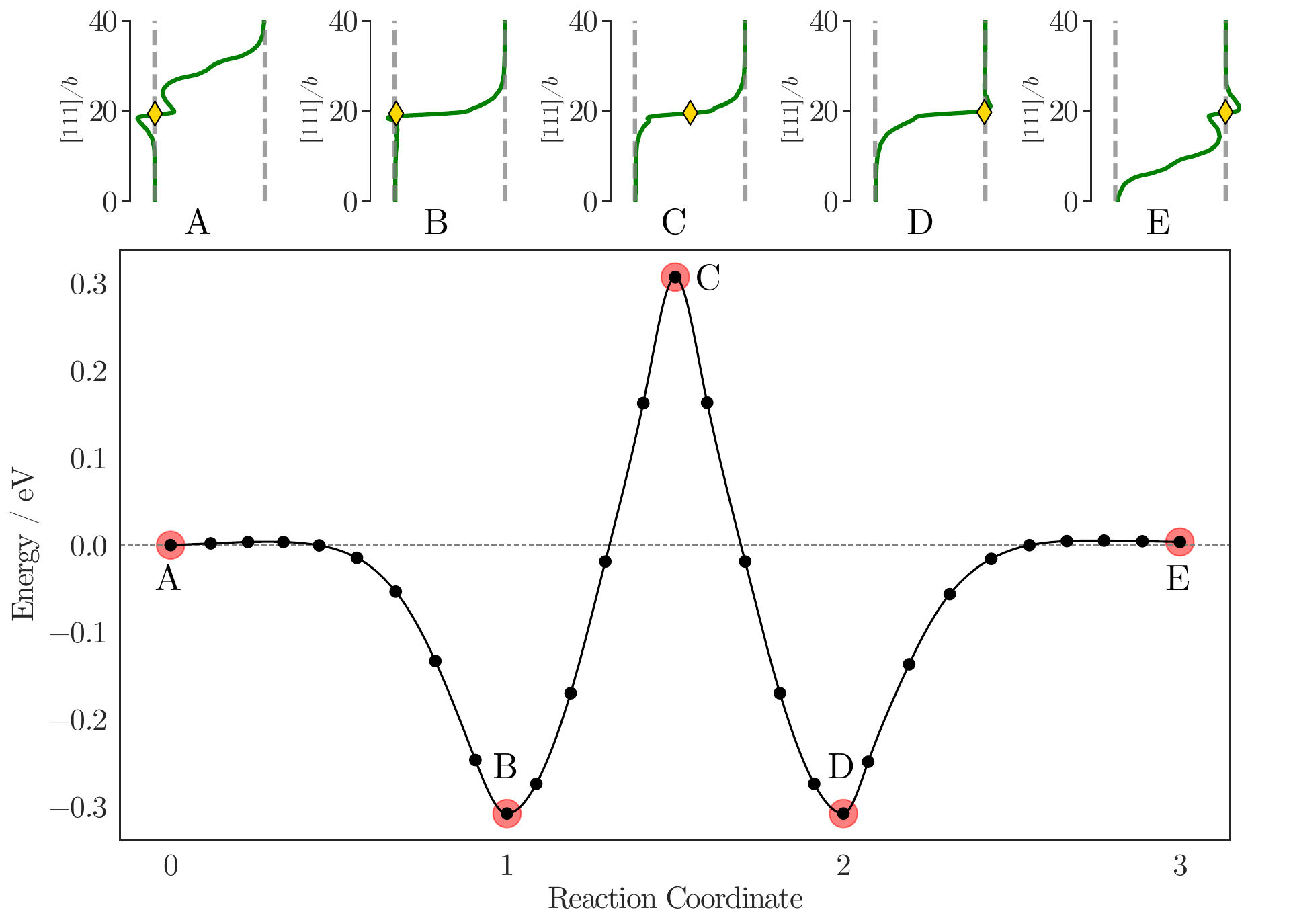}
        
        \textbf{(b)} Interstitial kink gliding past a helium interstitial.
    \end{minipage}

    \caption{Kink-pinning and unpinning pathways determined using the NEB method.
    A$\rightarrow$B: A free-gliding kink moves onto and becomes pinned by the helium atom.
    B$\rightarrow$D: Helium diffuses across the kink.
    D$\rightarrow$E: The kink is unpinned from the helium trap and resumes free glide.
    Reaction coordinates for A$\rightarrow$B and D$\rightarrow$E are based on the average dislocation position in the glide direction, calculated using the cost function.
    The reaction coordinate for B$\rightarrow$D is based on the helium position in the glide direction.}
    \label{fig:kink_unpinning}
\end{figure}

The mechanisms considered in the previous sections describe the single-cluster limit, in which the freely migrating kink does not encounter a second helium impurity. However, beyond this limit, a migrating kink may encounter an obstacle on the dislocation line. Figure~\ref{fig:kink_unpinning} shows the energy profiles for the migration of vacancy and interstitial kinks past an interstitial helium atom. 

Each pathway was constructed by joining three NEB image series describing kink pinning, helium diffusion across the kink, and kink unpinning. Helium diffusion across the kink allows the impurity to remain bound to the dislocation before and after the kink passes. In the initial and final configurations, the kink is separated from the helium atom by $12b$ along the dislocation line.

The kink pinning and unpinning pathways contain no intervening transition state: the energy decreases as the kink becomes pinned and rises as it moves away. The energy difference between the pinned and separated configurations therefore directly gives the unpinning barrier: 1.11~eV for the vacancy kink and 0.29~eV for the interstitial kink. Helium diffusion across the pinned kink has barriers of 22~meV and 0.62~eV, respectively. Consequently, unpinning presents the largest barrier to vacancy kink passage, whereas helium diffusion across the kink presents the largest barrier to interstitial kink passage.

Kink pinning provides a competing mechanism through which the mobility enhancement found in the single-cluster limit may be offset as the helium concentration increases. The stress dependence of these barriers remains to be established and could be investigated using an approach similar to that previously applied to silicon impurities in iron \cite{Shinzato2019-yt}.

\section{Conclusions}

In summary, using a new machine learning interatomic potential for dilute helium–dislocation interactions in tungsten, we have shown that individual helium atoms and small helium clusters bind substantially more strongly to screw dislocation kinks than to straight dislocation segments. This interaction is strongly asymmetric, with vacancy kinks providing considerably deeper binding sites than interstitial kinks. For a single helium atom, preferential stabilisation of the vacancy kink reduces the kink pair nucleation energy from 1.58~eV in pure tungsten to 0.48~eV.

For $\mathrm{He}_2$ and $\mathrm{He}_3$, the preference for binding to the vacancy kink exceeds the energy required to form a kink pair. Consequently, the kink pair configuration becomes the ground state, while the straight dislocation becomes a metastable intermediate along the minimum energy path for dislocation motion. The energy barrier for this increases from 0.60~eV for $\mathrm{He}_2$ to 1.00~eV for $\mathrm{He}_3$, indicating a progressive weakening of the enhanced mobility with increasing cluster size.

Molecular dynamics simulations at 900~K support the proposed mechanisms, showing enhanced screw dislocation mobility for $\mathrm{He}_1$ and $\mathrm{He}_2$. This softening effect becomes less pronounced with increasing cluster size and is not observed for $\mathrm{He}_3$ under the conditions studied. 

Although the potential retains close agreement with DFT for the configurations tested, the ground state ordering for the dislocation-$\mathrm{He}_2$ configuration and the quantitative energy barriers remain sensitive to errors in the underlying DFT training data and to details of the regression methodology. Because thermally activated rates depend exponentially on activation barriers, even small barrier errors can produce large errors in predicted rates.

These results challenge two simplifying assumptions used in some higher-scale models: that vacancy and interstitial kinks interact equivalently with impurities, and that binding to a straight dislocation can represent binding to kinked configurations \cite{Zhao2018-bn,Zhao2020-sm,Zhou2021-wm,Zhou2021-ws}. Both assumptions fail for W--He and should therefore be tested for each system. Explicitly distinguishing the two kink types and using kink-specific interaction energies should enable more physically grounded parameterisation of kinetic Monte Carlo and other mesoscale models of impurity-controlled dislocation motion.

More broadly, the mechanisms identified here show that dilute helium can enhance, rather than solely impede, screw dislocation motion in tungsten within the simulated single-cluster limit. The calculations presented here do not establish whether such a regime occurs during irradiation or how long it would persist. A transient softening regime is conceivable during the early stages of irradiation, before helium accumulation and the development of a more diverse defect population comprising larger helium clusters, self-interstitial atoms, and helium--vacancy complexes \cite{Zheng2021-bn}. Understanding how the mechanisms identified here evolve in this more realistic regime will be the focus of future work.

\section*{Data availability}
The \textit{ab initio} database, the ACE potential fitted to this database, and the atomistic configurations obtained from the NEB simulations that used the potential are available at \url{https://zenodo.org/records/18838361}.

\section*{Acknowledgments}

We thank James Turner and Shailesh Mehta for helpful discussions, and Fraser Birks for assistance with \texttt{ML-MIX}. MN acknowledges support from an EPSRC Postdoctoral Pathway Fellowship and a studentship funded by the EPSRC Centre for Doctoral Training in Modelling of Heterogeneous Systems, Grant No. EP/S022848/1 and AWE Nuclear Security Technologies. JRK acknowledges funding from the Leverhulme Trust under grant RPG-2017-191.
APB acknowledges support from the CASTEP-USER project, funded by the EPSRC under the grant agreement EP/W030438/1.
JRK and APB acknowledge funding from the NOMAD Centre of Excellence funded by the European Commission under grant agreement 951786.
We acknowledge the University of Warwick Scientific Computing Research Technology Platform for assisting the research described within this study. Some of the calculations were performed using the Sulis Tier 2 HPC platform hosted by the Scientific Computing Research Technology Platform at the University of Warwick. Sulis is funded by EPSRC Grant EP/T022108/1 and the HPC Midlands+ consortium. We are grateful for computational support from the UK national high performance computing service, ARCHER2, for which access was obtained via the UKCP consortium and funded by EPSRC Grant No. EP/X035891/1.

\clearpage

\appendix

\makeatletter
\@addtoreset{figure}{section}
\makeatother
\renewcommand{\thefigure}{\thesection\arabic{figure}}
\onecolumngrid

\section{Additional figures}
\label{app:additional_figures}

\begin{figure}[!htbp]
\centering

\includegraphics[width=0.85\linewidth]{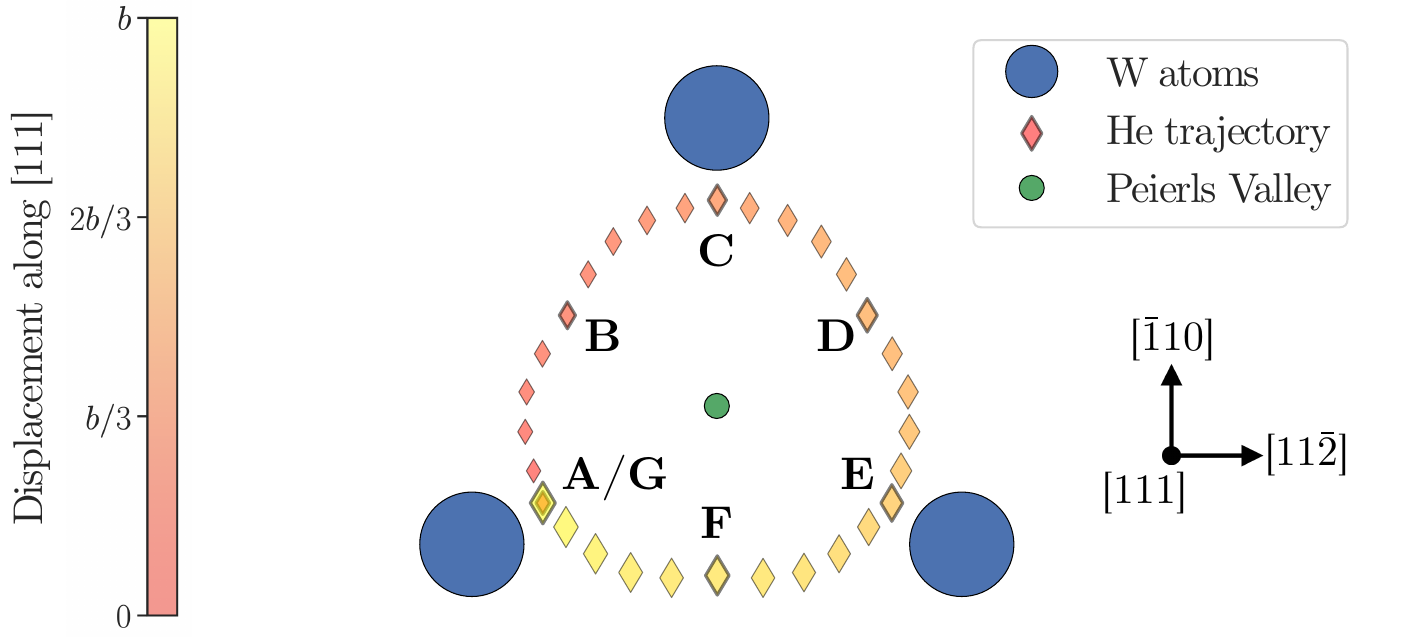}

\vspace{0.1cm}
{\raggedright
\small\textbf{(a)} Viewing down the dislocation line. The trajectory of the helium atom is shown and the position in [111] is indicated by colour and diamond size.\par
}

\vspace{0.3cm}

\includegraphics[width=0.85\linewidth]{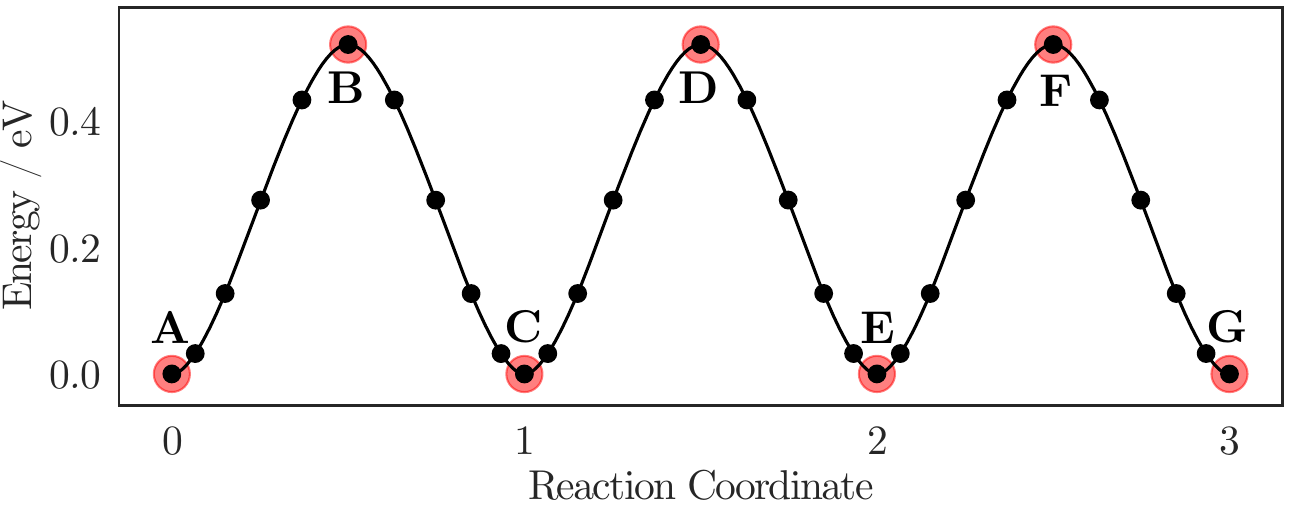}

\vspace{0.1cm}
{\raggedright
\small\textbf{(b)} Energy profile for the diffusion, where the reaction coordinate is based on the position of helium.\par
}

\vspace{0.3cm}

\includegraphics[width=0.85\linewidth]{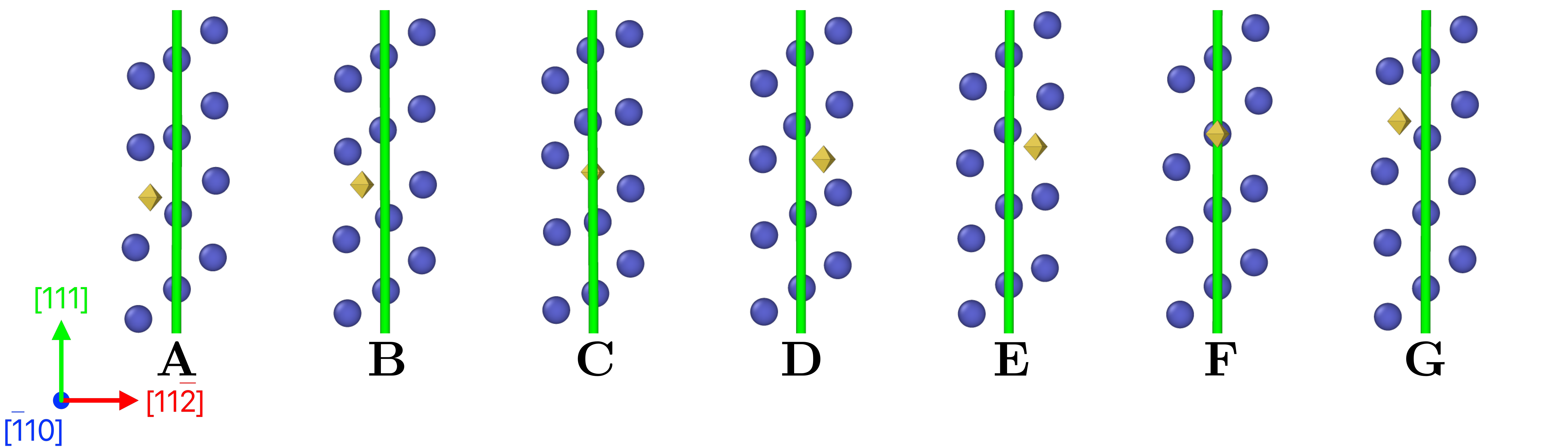}

\vspace{0.1cm}
{\raggedright
\small\textbf{(c)} Viewing along the line, image created with OVITO.\par
}

\caption{Diffusion of a helium interstitial along the dislocation line. After three equivalent hops, the helium has advanced along the line by $b$. Turning points of the energy profile are highlighted. Image series is generated from three NEB simulations.}
\label{fig:hediffhelix_combined}
\end{figure}

\begin{figure}
    \centering
    \includegraphics[width=0.95\linewidth]{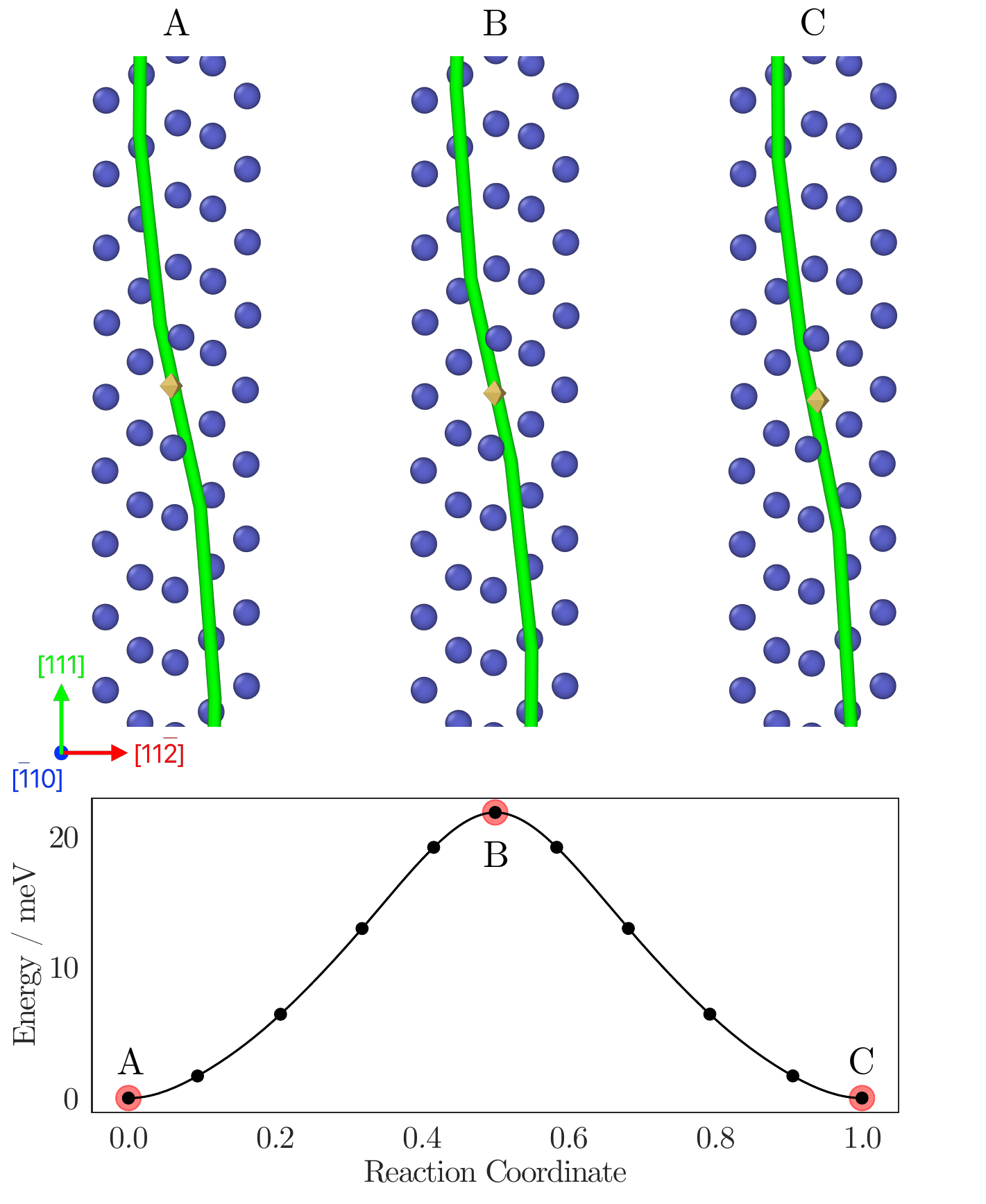}
    \caption{Diffusion of helium across the vacancy kink, simulated using the NEB method. Reaction coordinate is based on the position of helium in [11$\overline{2}$]. Atomistic configurations visualised with OVITO.}
    \label{fig:hediffrk}
\end{figure}
\begin{figure}
    \centering
    \includegraphics[width=0.95\linewidth]{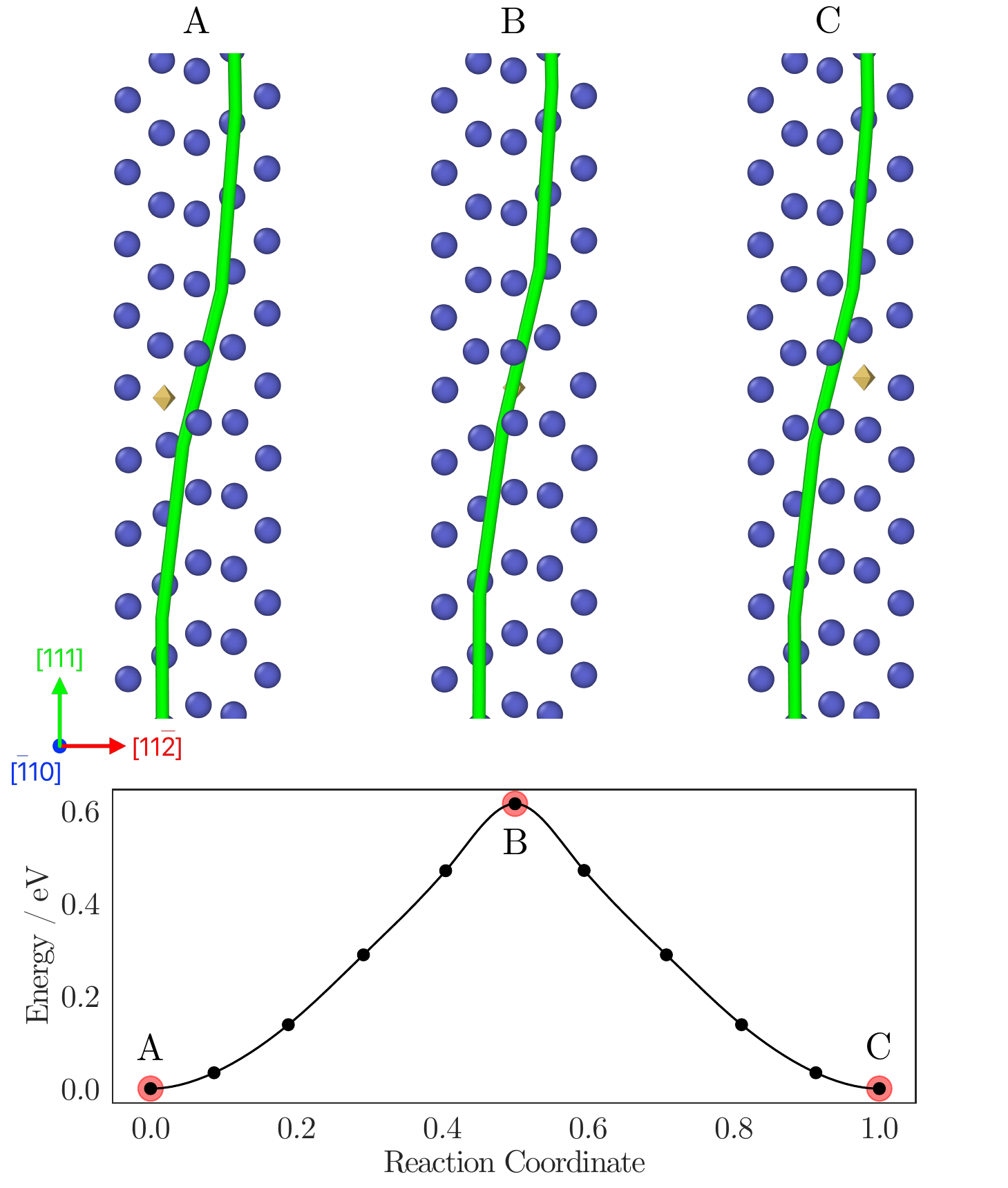}
    \caption{Diffusion of helium across the interstitial kink, simulated using the NEB method. Reaction coordinate is based on the position of helium in [11$\overline{2}$]. Atomistic configurations visualised with OVITO.}
    \label{fig:hedifflk}
\end{figure}

\pagebreak

\begin{figure}
    \centering
    \includegraphics[width=\textwidth]{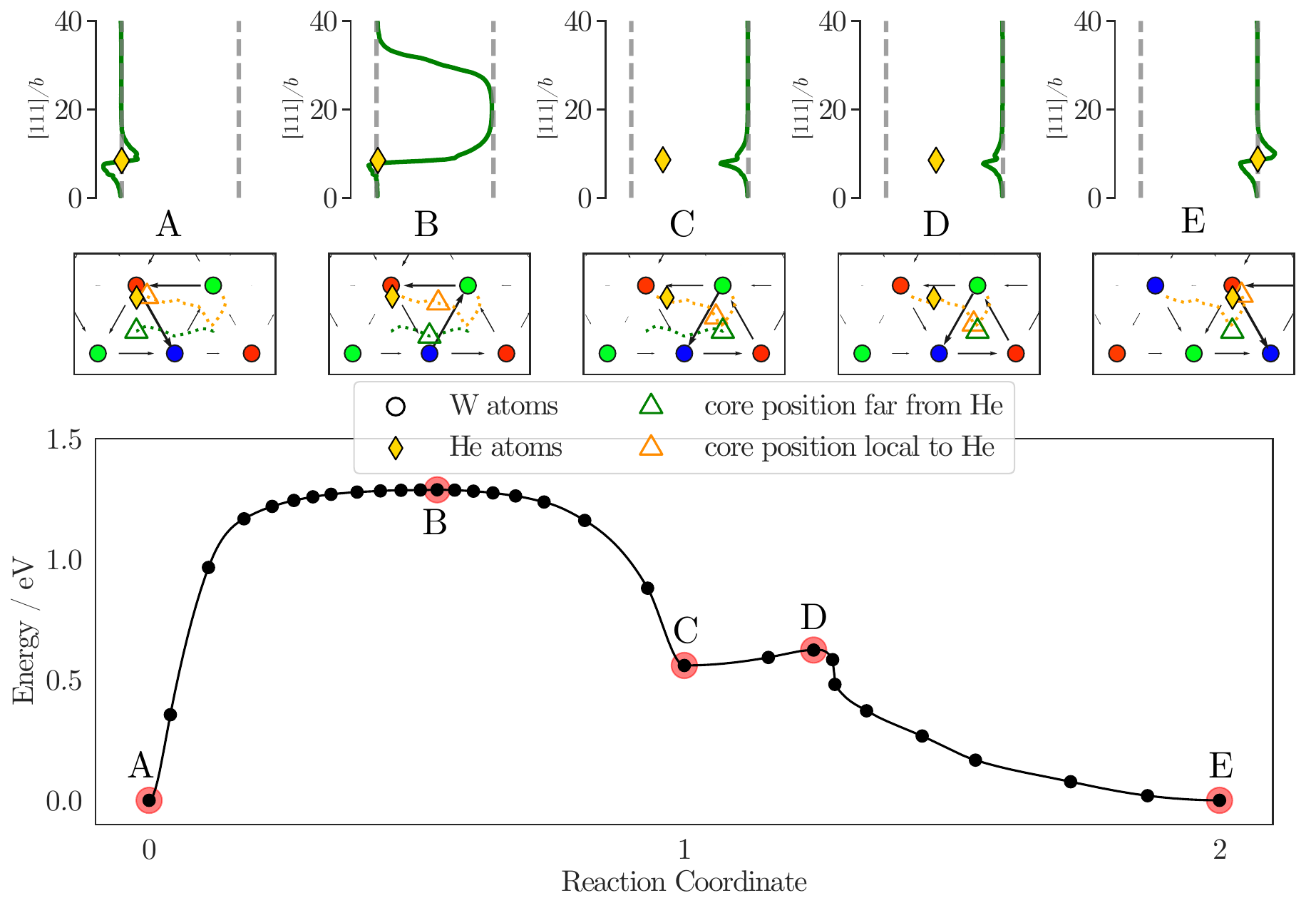}
    \caption{Two-step NEB pathway for helium-assisted kink pair nucleation, proceeding via the helium-stabilised interstitial kink. A$\rightarrow$C: Kink pair nucleation through helium-stabilised interstitial kink, with vacancy kink migration. C$\rightarrow$E: Diffusion of helium into the dislocation core. Dislocation line positions (determined using the cost function method \cite{Ventelon2013-zo}) and differential displacement maps for the layer local to helium for the key images shown (top and middle). In the Vitek maps, the colours of the [111] atomic columns change across the image series as the helium atom moves between layers. Green and orange dashed lines on the differential displacement maps represent the path taken by the dislocation far (20~\textit{b}) from and local to the helium, respectively. The triangles correspond to the position in the current image. Reaction coordinate is normalised by average dislocation position for A$\rightarrow$C and C$\rightarrow$E.}
    \label{fig:NEB1lk}
\end{figure}

\twocolumngrid

\bibliography{paperpile}
\bibliographystyle{apsrev4-1}

\end{document}